\documentclass[11pt]{article}
\usepackage{amssymb}

\input{tcilatex}
\begin{document}

\author{J. G. Cardoso\thanks{%
jorge.cardoso@udesc.br} \\
Department of Mathematics\\
Centre for Technological Sciences-UDESC\\
Joinville 89219-710 SC\\
Brazil.}
\title{ A Torsional Two-Component Description of the Motion of Dirac
Particles at Early Stages of the Cosmic Evolution}
\date{ }
\maketitle

\begin{abstract}
It is assumed that the non-singular big-bang birth of the Universe as set
forth by Einstein-Cartan's theory particularly brought about the appearance
of the cosmic microwave and dark energy backgrounds, dark matter, gravitons
as well as of Dirac particles. On account of this assumption, a
two-component description of the motion of quarks and leptons prior to the
occurrence of hadronization is presented within the framework of the
torsionful $\varepsilon $-formalism of Infeld and van der Waerden. The
relevant field equations are settled on the basis of the implementation of
conjugate minimal coupling covariant derivative operators that carry
additively typical potentials for the cosmic backgrounds such as
geometrically specified in a previous work. It appears that the derivation
of the wave equations which control the spacetime propagation of Dirac
fields at very early stages of the cosmic evolution, must be tied up with
the applicability of certain subsidiary relations. The wave equations
themselves suggest that quarks and leptons interact not only with both of
the cosmic backgrounds, but also with dark matter. Nevertheless, it becomes
manifest that the inner structure of the framework allowed for does not give
rise at all to any interaction between gravitons and Dirac particles. The
overall formulation ascribes an intrinsically non-geometric character to
Dirac's theory, in addition to exhibiting a formal evidence that dark energy
and dark matter must have partaken of a cosmic process of hadronization.
\end{abstract}

\section{Introduction}

An important physical property of Einstein-Cartan's theory [1-3] is related
to the fact that it yields torsional cosmological models which describe a
non-singular big-bang birth of the Universe [4,5]. This feature is seemingly
in contrast to the context of Einstein's theory of general relativity, which
gives rise to isotropic and homogeneous cosmological models that predict the
occurrence of unavoidably singular gravitational collapses [6-11]. Because
of the characteristic asymmetry borne by the Ricci tensor for any torsionful
world affine connexion, the right-hand sides of Einstein-Cartan's field
equations always carry asymmetric energy-momentum tensors. The skew parts of
such tensors were originally identified [12-14] with sources for densities
of intrinsic angular momentum of matter which presumably generate spacetime
torsion locally. What has indeed ensured the consistency of the construction
of the aforesaid torsional cosmological models, which are sometimes
designated as Trautman-Kopczy\'{n}ski cosmological models, stems from the
establishment [15] that Einstein-Cartan's equations admit a two-parameter
family of spherically symmetric solutions of the Friedmann type, which
supply a lower bound for the final radius of a gravitationally collapsing
cloud of dust. Thus, according to these torsional models, the Universe has
expanded from a spherically symmetric state having a finite radius. One of
the central results arising in Ref. [15] is that the torsional cosmological
models of the Universe must not be spatially homogeneous inasmuch as the
classical Friedmann homogeneity property gets lost when torsion is brought
into the spacetime geometry. This result apparently exhibits a contextual
compatibility with the work of Ref. [16] which shows explicitly that if the
Friedmann cosmological principle such as stated classically is allowed for
within the realm of Einstein-Cartan's theory, then all the components of the
torsion tensor shall turn out to vanish identically. Furthermore, it has
really provided us with the most significant physical meaning of the
correspondence principle that interrelates Einstein-Cartan's theory and
General Relativity. Based upon a concept of microscopically produced
spacetime torsion, the work of Ref. [17] has proposed some torsional
mechanisms that solve naturally the cosmological spatial flatness and
horizon problems, whilst showing that Dirac particles may have generated a
small amount of torsion immediately after the occurrence of the big-bang,
which could contribute to a gravitational repulsion that prevents the
cosmological singularity to occur.

Amongst the elementary spacetime properties of Einstein-Cartan's theory,
there is a striking feature of this theory which bears a deeper mathematical
character, namely, the fact that any spacetime endowed with a torsionful
affinity admits a local spinor structure in much the same way as in the case
of generally relativistic spacetimes [18,19]. This admissibility has
supported the construction of an essentially unique torsional extension [20]
of the famous two-component spinor $\gamma \varepsilon $-formalisms of
Infeld and van der Waerden for general relativity [21-23], which was
primarily aimed at proposing a local description whereby dark energy should
be presumptively looked upon as a torsional cosmic background [24]. Dark
energy fields were thus defined as uncharged spin-one massive fields which
come from contracted spin-affine pieces that amount to gauge invariant Proca
potentials. Roughly speaking, such invariant spin-affine contributions
account for the underlying spacetime torsionfulness, and thereby also enter
additively together with contracted spin-affine pieces for the cosmic
microwave background into overall contracted spin affinities. A notable
result arising in this framework is that, in contradistinction to dark
energy, the cosmic microwave background may propagate alone in torsionless
regions of the Universe. It was realized, in effect, that any torsional
affine potential must be accompanied by a suitable torsionless spin-affine
contribution in such a way that the propagation of the dark energy
background has to be geometrically united together with that of the cosmic
microwave background, whence the cosmic backgrounds had to come about
together during the birth of the Universe. However, it had become manifest
thereabout that one background may propagate in spacetime as if the other
were absent, such that they should not be taken to interact with one
another. Accordingly, it turns out that the propagation of the cosmic
microwave background in regions of the Universe where the values of torsion
are negligible, can be described alone within the torsionless $\gamma
\varepsilon $-framework.

More recently, it has been suggested [25] that the torsionful extension of
the generally relativistic $\gamma \varepsilon $-formalisms referred to
before, should supply a physical description of wave functions for dark
matter in terms of uncharged spin-one massive fields which were supposedly
produced together with gravitons. This suggestion stemmed from the belief
that wave functions for dark matter arise together with ones for gravitons
in the irreducible expansion for one of the Witten curvature spinors of a
torsionful spin affinity [20, 26]. It was then pointed out that a geometric
formulation of the dynamics of dark matter might be accomplished out of the
combination of Einstein-Cartan's theory with the torsionful $\gamma
\varepsilon $-framework, which would possibly explain some of the currently
observed features of the Universe. Likewise, a reasonable spinor version of
Einstein-Cartan's equations [25] has exhibited a purely microscopic
identification as dark matter of the densities of spinning matter that are
theoretically associated to the local production of spacetime torsion.

In the present paper, we allow for the view that the non-singular big-bang
creation of the Universe as described by the cosmological models of
Trautman-Kopczy\'{n}ski [4,5,15], had produced a highly torsional hot state
which brought about from the beginning the appearance of the cosmic
microwave and dark energy backgrounds, dark matter, gravitons as well as of
Dirac particles. A two-component description of the motion of quarks and
leptons prior to the occurrence of hadronization is then formulated within
the framework of the torsionful $\varepsilon $-formalism. The relevant field
equations are settled on the basis of the implementation of conjugate
minimal coupling covariant derivative operators that appropriately carry
affine potentials for the cosmic backgrounds such as geometrically specified
in Ref. [24]. It thus appears that the derivation of the wave equations
which control the spacetime propagation of Dirac fields at very early stages
of the cosmic evolution, must be bound up with the appositeness of
subsidiary relations that take up typical Maxwell and Proca potentials, with
the equations themselves suggesting that quarks and leptons interact not
only with both of the cosmic backgrounds, but also with dark matter.
Nonetheless, we will see that the inner structure of the framework to be
taken into consideration herein does not give rise at all to any interaction
between gravitons and Dirac particles. It shall be seen that the overall
formulation ascribes an intrinsically non-geometric character to Dirac's
theory, in addition to exhibiting a formal evidence that dark energy and
dark matter must have partaken of a cosmic process of hadronization.

All the world and spinor conventions adhered to in Ref. [20], will be
adopted throughout the work. In particular, vertical bars surrounding an
index block will mean that the indices singled out should not be absorbed by
a symmetry operation, while symmetrizations and skew-symmetrizations over
index blocks will be denoted by round and square brackets surrounding the
indices sorted out by the symmetry operations, respectively. For
organizational reasons, we will have to recall in Sections 2 and 3 the world
version of Einstein-Cartan's theory along with a brief review of the works
of Refs. [24,25]. Upon developing Section 3, we shall take for granted the
same calculational techniques as the ones built up in Ref. [24], but these
techniques will be alluded to in Section 4. Insofar as the $\varepsilon $%
-formalism version of Dirac wave functions must be identified with
spin-vector densities of adequate weights, some fragments of the theory of
geometric spin densities for the generalized Weyl gauge group [20,22] shall
be allowed for at the outset. It will become clear that the only need for
placing minimal coupling covariant derivative operators into the field
dynamics, relies crucially on this identification. The presentation of
Section 3 will make it feasible to draw up some remarks in Section 5, which
are related to our broader interest in bringing forward two-component
patterns for earlier cosmic interaction couplings that involve Dirac
particles. We will use the natural system of units wherein $c=\hslash =1.$

\section{Spacetime Torsion and Einstein-Cartan's Theory}

In the context of Einstein-Cartan's theory, any spacetime carries a
symmetric metric tensor $g_{\mu \nu }$ together with a torsionful covariant
derivative operator $\nabla _{\mu }$ that fulfills the metric compatibility
condition $\nabla _{\mu }g_{\lambda \sigma }=0.$ For convenience, $g_{\mu
\nu }$ is usually required to bear the local signature $(+---).$ The Riemann
tensor $R_{\mu \nu \lambda }{}^{\sigma }$ of $\nabla _{\mu }$ occurs in
either of the expansions%
\begin{equation}
D_{\mu \nu }u^{\alpha ...\beta }=R_{\mu \nu \tau }{}^{\alpha }u^{\tau
...\beta }{}+\cdots +R_{\mu \nu \tau }{}^{\beta }u^{\alpha ...\tau }{}
\label{1}
\end{equation}%
and%
\begin{equation}
D_{\mu \nu }w_{\lambda ...\sigma }=-R_{\mu \nu \lambda }{}^{\tau }w_{\tau
...\sigma }-\cdots -R_{\mu \nu \sigma }{}^{\tau }w_{\lambda ...\tau },
\label{2}
\end{equation}%
where $u^{\alpha ...\beta }{}$ and $w_{\lambda ...\sigma }$ are some world
tensors, and%
\begin{equation}
D_{\mu \nu }=2(\nabla _{\lbrack \mu }\nabla _{\nu ]}+T_{\mu \nu }{}^{\lambda
}\nabla _{\lambda }).  \label{3}
\end{equation}%
Hence, the operator $D_{\mu \nu }$ obeys the Leibniz rule. When acting on a
world-spin scalar $h,$ it gives $D_{\mu \nu }h=0$ by definition. The object $%
T_{\mu \nu }{}^{\lambda }$ is the torsion tensor of $\nabla _{\mu }.$ It
just amounts to the skew part $\Gamma _{\lbrack \mu \nu ]}{}^{\lambda }$ of
the world affinity $\Gamma _{\mu \nu }{}^{\lambda }$ of $\nabla _{\mu }$
and, then, satisfies the relation%
\begin{equation}
T_{\mu \nu }{}^{\lambda }=T_{[\mu \nu ]}{}^{\lambda },  \label{4}
\end{equation}%
which thus yields the property $D_{\mu \nu }=D_{[\mu \nu ]}.$ For the
corresponding Ricci tensor and scalar, one has%
\begin{equation}
R_{\mu \nu }=R_{\mu \lambda \nu }{}^{\lambda },\text{ }R=R^{\lambda \sigma
}g_{\lambda \sigma }.  \label{5}
\end{equation}

The tensor $R_{\mu \nu \lambda \sigma }{}$ possesses skewness in the indices
of the pairs $\mu \nu $ and $\lambda \sigma ,$ but the\ traditional
index-pair symmetry $R_{\mu \nu \lambda \sigma }=R_{\lambda \sigma \mu \nu }$
ceases holding because of the applicability of the cyclic identity [20]%
\begin{equation}
^{\ast }R^{\lambda }{}_{\mu \nu \lambda }+2\nabla ^{\lambda }{}^{\ast
}T_{\lambda \mu \nu }+4{}^{\ast }T_{\mu }{}^{\lambda \tau }{}T_{\lambda \tau
\nu }=0,  \label{6}
\end{equation}%
where%
\begin{equation}
^{\ast }R{}_{\mu \nu \lambda \sigma }=\frac{1}{2}\sqrt{-\mathfrak{g}}e_{\mu
\nu \rho \tau }R^{\rho \tau }{}_{\lambda \sigma }  \label{nLinLin}
\end{equation}%
and%
\begin{equation}
^{\ast }T_{\mu \nu }{}^{\lambda }=\frac{1}{2}\sqrt{-\mathfrak{g}}e_{\mu \nu
\rho \tau }T^{\rho \tau \lambda },  \label{nLinLin1}
\end{equation}%
define first-left duals [3], with $\mathfrak{g}$ accordingly denoting the
determinant of $g_{\mu \nu }$ and $e_{\mu \nu \rho \tau }$ standing for one
of the invariant world Levi-Civitta densities. Thus, the Ricci tensor of $%
\nabla _{\mu }$ bears asymmetry whilst the gravitational Bianchi identity
should now read%
\begin{equation}
\nabla ^{\rho }{}^{\ast }R_{\rho \mu \lambda \sigma }+2{}^{\ast }T{}{}_{\mu
}{}^{\rho \tau }R_{\rho \tau \lambda \sigma }=0.  \label{7}
\end{equation}%
\qquad

From the metric compatibility condition for $\nabla _{\mu },$ we can deduce
the equivalent relations [24]%
\begin{equation}
\widetilde{\nabla }_{\mu }g_{\lambda \sigma }-2T_{\mu (\lambda \sigma )}=0
\label{n2}
\end{equation}%
and%
\begin{equation}
\Gamma _{\mu }=\widetilde{\Gamma }_{\mu }+T_{\mu }=\partial _{\mu }\log (-%
\mathfrak{g})^{1/2},  \label{n2Lin}
\end{equation}%
where $T_{\mu }=T_{\mu \lambda }{}^{\lambda },$ for instance, and%
\begin{equation}
\Gamma _{(\mu \lambda )\sigma }\doteqdot \widetilde{\Gamma }_{\mu \lambda
\sigma },  \label{n9}
\end{equation}%
with $\widetilde{\nabla }_{\mu }$ amounting to the covariant derivative
operator for $\widetilde{\Gamma }_{\mu \lambda \sigma }.$ Thus, putting into
effect the secondary metric condition%
\begin{equation}
\widetilde{\nabla }_{\lambda }g_{\mu \nu }=0,  \label{addLin1}
\end{equation}%
implies that $T_{\mu }=0$ everywhere in spacetime and, consequently,%
\begin{equation}
T_{\mu \nu \lambda }=T_{[\mu \nu \lambda ]},  \label{in2}
\end{equation}%
whence $T_{\mu \nu \lambda }$ will turn out to possess only four independent
components in case the condition (\ref{addLin1}) is actually implemented.
Under this affine circumstance, one may recover the characteristic 24
torsional components by making use of the contortion tensor for $T_{\mu \nu
\lambda }$ while identifying $\widetilde{\Gamma }_{\mu \lambda \sigma }$
with a Christoffel connexion.

The tensor $g_{\mu \nu }$ comes into play as a solution to Einstein-Cartan's
equations\footnote{%
The quantity $\kappa $ is still identified with Einstein's gravitational
constant of general relativity.}%
\begin{equation}
R_{\mu \nu }-\frac{1}{2}Rg_{\mu \nu }=-\kappa E_{\mu \nu },  \label{8}
\end{equation}%
whilst $T_{\mu \nu }{}^{\lambda }$ is locally related [3] to the spin
density of matter $S_{\mu \nu }{}^{\lambda }$ present in spacetime through%
\begin{equation}
T_{\mu \nu }{}^{\lambda }=-\kappa (S_{\mu \nu }{}^{\lambda }-S_{[\mu }g_{\nu
]}{}^{\lambda }),  \label{9}
\end{equation}%
with $S_{\mu }\doteqdot S_{\mu \sigma }{}^{\sigma }.$ Equation (\ref{9})
leads right away to the trace relation 
\begin{equation}
T_{\mu }{}=\frac{\kappa }{2}S_{\mu },  \label{10}
\end{equation}%
whence we can write down the supplementary relationship%
\begin{equation}
-\kappa S_{\mu \nu }{}^{\lambda }=T_{\mu \nu }{}^{\lambda }-2T_{[\mu }g_{\nu
]}{}^{\lambda }.  \label{11}
\end{equation}%
It follows that, working out the relation (\ref{6}), produces the equation
[25]%
\begin{equation}
\nabla _{\lambda }T_{\mu \nu }{}^{\lambda }+2(\nabla _{\lbrack \mu }T_{\nu
]}{}+T_{\mu \nu }{}^{\lambda }T_{\lambda })=\kappa E_{[\mu \nu ]},
\label{12}
\end{equation}%
which amounts to the same thing as%
\begin{equation}
(\nabla _{\lambda }+\kappa S_{\lambda })S_{\mu \nu }{}^{\lambda }=-E_{[\mu
\nu ]}.  \label{13}
\end{equation}%
It can therefore be said that the skew part of $E_{\mu \nu }$ is a source
for $S_{\mu \nu }{}^{\lambda },$ and thence also for $T_{\mu \nu
}{}^{\lambda }.$ Hence, setting $T_{\mu }=0,$ leads to the reduced equations%
\footnote{%
Equations (\ref{ADD1}) and (\ref{ADD2}) play a significant role in the work
of Ref. [25] in that they are involved in the proposal of a simpler
expression for the mass of dark matter.}%
\begin{equation}
\nabla _{\lambda }T_{\mu \nu }{}^{\lambda }=\kappa E_{[\mu \nu ]},\text{ }%
\nabla _{\lambda }S_{\mu \nu }{}^{\lambda }=-E_{[\mu \nu ]},  \label{ADD1}
\end{equation}%
which obviously agree with the particular relation%
\begin{equation}
T_{\mu \nu }{}^{\lambda }=-\kappa S_{\mu \nu }{}^{\lambda }.  \label{ADD2}
\end{equation}

\section{Cosmic Backgrounds and Dark Matter}

We shall now review briefly the descriptions of the cosmic backgrounds and
dark matter as suggested in Refs. [24,25]. This procedure will afford a
necessary basis for the settlement of Sections 4 and 5.

The key curvature object for the $\varepsilon $-formalism is a gauge
invariant spin-tensor density $C_{\mu \nu CD}{}$ of weight $-1.$ By
definition, it occurs in the configuration [20]%
\begin{equation}
D_{\mu \nu }\alpha ^{D}=C_{\mu \nu M}{}^{D}\alpha ^{M},  \label{15}
\end{equation}%
where $\alpha ^{D}$ amounts to an arbitrary spin vector and $D_{\mu \nu }$
is given by Eq. (\ref{3}). The contracted piece $C_{\mu \nu M}{}^{M}{}$
possesses the additivity property%
\begin{equation}
C_{\mu \nu M}{}^{M}=-2i(\widetilde{F}_{\mu \nu }+F_{\mu \nu }^{%
{\footnotesize (T)}}),  \label{16}
\end{equation}%
where%
\begin{equation}
\widetilde{F}_{\mu \nu }\doteqdot 2\partial _{\lbrack \mu }\Phi _{\nu ]},%
\text{ }F_{\mu \nu }^{{\footnotesize (T)}}\doteqdot 2\partial _{\lbrack \mu
}A_{\nu ]},  \label{17}
\end{equation}%
with $\Phi _{\mu }$ and $A_{\mu }$ being affine potentials subject to the
gauge behaviours\footnote{%
The quantity $\Theta $ stands for the Weyl gauge parameter. In Ref.[22],
this parameter is shown to be a world-spin invariant such that $\partial
_{\mu }\Theta =\nabla _{\mu }\Theta .$}%
\begin{equation}
\Phi _{\mu }^{\prime }=\Phi _{\mu }-\partial _{\mu }\Theta ,\text{ }A_{\mu
}^{\prime }=A_{\mu }.  \label{18}
\end{equation}%
In the description proposed in Ref. [24], $\Phi _{\mu }$ is assumably taken
as a potential for the cosmic microwave background, while $A_{\mu }$ is
considered thereabout as a Proca potential for the dark energy background.
The world form of the first half of the overall set of field equations,
appears as a combination of the first half of Maxwell's equations 
\begin{equation}
\nabla ^{\mu }\widetilde{F}_{\mu \lambda }+2T^{\mu }\widetilde{F}_{\mu
\lambda }-T{}^{\mu \nu }{}_{\lambda }\widetilde{F}_{\mu \nu }=0,\text{ }
\label{20}
\end{equation}%
with the first half of Proca's equations%
\begin{equation}
\nabla ^{\mu }F_{\mu \lambda }^{{\footnotesize (T)}}+2T^{\mu }F_{\mu \lambda
}^{{\footnotesize (T)}}-T{}^{\mu \nu }{}_{\lambda }F_{\mu \nu }^{%
{\footnotesize (T)}}+m^{2}A_{\lambda }=0,  \label{21}
\end{equation}%
where $m$ is the mass of $A_{\mu }.$ Both of the second halves come forth as
the corresponding Bianchi identities, which may be symbolically expressed as
the dual statement%
\begin{equation}
\nabla ^{\mu }{}^{\ast }\mathcal{F}_{\mu \lambda }=-2{}^{\ast }T_{\lambda
}{}^{\mu \nu }\mathcal{F}_{\mu \nu },  \label{23}
\end{equation}%
with the kernel letter $\mathcal{F}$ standing for either $\widetilde{F}$ or $%
F^{{\footnotesize (T)}}.$

The spinor decompositions of the bivectors carried by Eq. (\ref{16}) thus
provide wave functions for the cosmic backgrounds, according to the scheme%
\begin{equation}
\widetilde{F}_{\mu \nu }\mapsto (\phi _{AB},\phi _{A^{\prime }B^{\prime }}),%
\text{ }F^{{\footnotesize (T)}}\mapsto (\psi _{AB},\psi _{A^{\prime
}B^{\prime }}),  \label{25}
\end{equation}%
where the entries of the pairs $(\phi _{AB},\phi _{A^{\prime }B^{\prime }})$
and $(\psi _{AB},\psi _{A^{\prime }B^{\prime }})$ amount to symmetric
massless and massive spin-one uncharged wave functions of opposite
handednesses, respectively, which obey the field-potential relations (see
Eq. (\ref{tauone}) below)%
\begin{equation}
\phi _{AB}{}=-\nabla _{(A}^{C^{\prime }}\Phi _{B)C^{\prime }}+2\tau
_{AB}{}^{\mu }\Phi _{\mu }  \label{26}
\end{equation}%
and%
\begin{equation}
\psi _{AB}{}=-\nabla _{(A}^{C^{\prime }}A_{B)C^{\prime }}+2\tau _{AB}{}^{\mu
}A_{\mu }.  \label{26Lin}
\end{equation}%
The pertinent spinor field equations arise when the transcription of Eqs. (%
\ref{20})-(\ref{23}) is carried through. For the unprimed wave functions,
say, we then have (for full details, see Ref. [24])%
\begin{equation}
\nabla _{B^{\prime }}^{A}\phi _{AB}+\mathcal{M}_{BB^{\prime }}=0  \label{27}
\end{equation}%
and%
\begin{equation}
\nabla _{B^{\prime }}^{A}\psi _{AB}+M_{BB^{\prime }}+\frac{1}{2}%
m^{2}A_{BB^{\prime }}=0,  \label{27Lin}
\end{equation}%
along with the individual terms%
\begin{equation}
\mathcal{M}_{BB^{\prime }}=2(T_{B^{\prime }}^{A}\phi _{AB}-\tau
^{AM}{}_{BB^{\prime }}\phi _{AM})  \label{29}
\end{equation}%
and%
\begin{equation}
M_{BB^{\prime }}=2(T_{B^{\prime }}^{A}\psi _{AB}-\tau ^{AM}{}_{BB^{\prime
}}\psi _{AM}),  \label{30}
\end{equation}%
with the $\tau T$-factors of (\ref{29}) and (\ref{30}) entering the
world-spinor relationships%
\begin{equation}
T_{\lambda \sigma }{}^{\mu }\mapsto (\tau _{AB}{}^{\mu },\tau _{A^{\prime
}B^{\prime }}{}^{\mu }),\text{ }T_{\mu }\mapsto T_{AA^{\prime }}.
\label{tauone}
\end{equation}

To derive the wave equations that control the propagation in spacetime of
the cosmic backgrounds [24], one has to call for the gravitational
correspondence%
\begin{equation}
R_{\mu \nu \lambda \sigma }\mapsto (\text{X}_{ABCD},\Xi _{A^{\prime
}B^{\prime }CD}),  \label{31}
\end{equation}%
with the X$\Xi $-spinors showing up as the Witten curvature spinors for $%
\nabla _{\mu }.$ In this connection, one must appeal to the symbolic
algebraic and valence-reduction devices given in Ref. [3], to obtain the
expansion [20]%
\begin{equation}
\text{X}_{ABCD}\hspace{-0.07cm}=\hspace{-0.07cm}\Psi _{ABCD}-\varepsilon
_{(A\mid (C}\xi _{D)\mid B)}-\frac{1}{3}\varkappa \varepsilon
_{A(C}\varepsilon _{D)B},  \label{add6}
\end{equation}%
together with the definitions\footnote{%
The objects $(\varepsilon _{AB},\varepsilon _{A^{\prime }B^{\prime }})$ are
the only covariant metric spinors for the formalism being allowed for [22],
and satisfy covariant constancy requirements of the type $\nabla _{\mu
}\varepsilon _{AB}=0.$}%
\begin{equation}
\hspace{-0.07cm}\Psi _{ABCD}=\text{X}_{(ABCD)}\hspace{-0.07cm},\text{ }\xi
_{AB}=\text{X}^{M}{}_{(AB)M},\text{ }\varkappa =\text{X}_{LM}{}^{LM},
\label{add7}
\end{equation}%
with the $\Psi $-spinor standing for a typical wave function for gravitons,
and $\varkappa $ being a complex-valued world-spin invariant. It was
suggested in Ref. [25] that the gravitational interaction between the $\phi
\psi $-fields and torsion, such as prescribed by Eqs. (\ref{29}) and (\ref%
{30}), generates a partial amount of mass for each of the cosmic
backgrounds. When the condition (\ref{addLin1}) is effectively accounted
for, the information on the whole acquired masses may be supplied from the
explicit derivatives carried by the Klein-Gordon patterns%
\begin{equation}
\mu _{CMB}^{2}\phi _{AB}=-4\nabla _{\mu }(\phi _{C(A}\tau _{B)}^{C}{}^{\mu
})-2\nabla _{(A}^{C^{\prime }}\mathcal{M}_{B)C^{\prime }}  \label{d1}
\end{equation}%
and\footnote{%
The work of Ref. [25] shows us that the relation $\nabla _{\mu }\tau
_{AB}{}^{\mu }=\xi _{AB}$ holds whenever the condition (\ref{addLin1}) is
implemented. This relation constitutes the spinor version of the first of
Eqs. (\ref{ADD1}).}%
\begin{equation}
\mu _{DE}^{2}\psi _{AB}=-4\nabla _{\mu }(\psi _{C(A}\tau _{B)}^{C}{}^{\mu
})-2\nabla _{(A}^{C^{\prime }}M_{B)C^{\prime }}.  \label{d2}
\end{equation}%
Provided that, for example,%
\begin{equation}
(\nabla _{\mu }\psi _{C(A})\tau _{B)}^{C\mu }=\nabla _{\mu }(\psi _{C(A}\tau
_{B)}^{C\mu })-\xi _{(A}^{C}\psi _{B)C},  \label{add100}
\end{equation}%
the $\varepsilon $-formalism wave equations for the cosmic backgrounds turn
out to be set up as the invariant statements [24,25]%
\begin{equation}
(\square +\mu _{CMB}^{2}+\frac{4}{3}\varkappa )\phi _{AB}-2\Psi
{}_{AB}{}^{CD}\phi _{CD}+2\xi _{(A}^{C}\phi _{B)C}=0  \label{N6}
\end{equation}%
and%
\begin{equation}
(\square +\mu _{eff}^{2}+\frac{4}{3}\varkappa )\psi _{AB}-2\Psi
{}_{AB}{}^{CD}\psi _{CD}+2\xi _{(A}^{C}\psi _{B)C}=2m^{2}\tau _{AB}{}^{\mu
}{}A_{\mu },  \label{N7}
\end{equation}%
with the effective-mass term%
\begin{equation}
\mu _{eff}=\sqrt{m^{2}+\mu _{DE}^{2}}.  \label{N2}
\end{equation}

The principal attitude taken in Ref. [25] involves identifying the world
field equation for dark matter as the following version of the Bianchi
identity (\ref{7}) 
\begin{equation}
\nabla ^{\rho \text{ }\ast }R^{\lambda }{}_{[\rho \sigma ]\lambda }=-2^{\ast
}T^{\lambda \rho \tau }R_{\tau \lbrack \rho \sigma ]\lambda }.  \label{61}
\end{equation}%
By utilizing the relationship\footnote{%
The symbol "c.c." denotes here as elsewhere an overall complex conjugate
piece}%
\begin{equation}
^{\ast }R^{\lambda }{}_{[\rho \sigma ]\lambda }\mapsto {}^{\ast
}R^{DD^{\prime }}{}_{(BC)[B^{\prime }C^{\prime }]DD^{\prime }}+\text{c.c.},
\label{62}
\end{equation}%
along with the expression [20]%
\begin{equation}
^{\ast }R_{AA^{\prime }BB^{\prime }CC^{\prime }DD^{\prime
}}=[(-i)(\varepsilon _{A^{\prime }B^{\prime }}\varepsilon _{C^{\prime
}D^{\prime }}\text{X}_{ABCD}-\varepsilon _{AB}\varepsilon _{C^{\prime
}D^{\prime }}\Xi _{A^{\prime }B^{\prime }CD})]+\text{c.c.},  \label{R2}
\end{equation}%
after invoking Eq. (\ref{add6}), one gets the uncharged decomposition%
\begin{equation}
^{\ast }R^{DD^{\prime }}{}_{(BC)[B^{\prime }C^{\prime }]DD^{\prime }}+\text{%
c.c.}=-i(\varepsilon _{B^{\prime }C^{\prime }}\xi _{BC}{}-\text{c.c.}).
\label{63}
\end{equation}%
The entries of the pair $(\xi _{AB},\xi _{A^{\prime }B^{\prime }})$ were
then thought of as constituting wave functions of opposite handednesses for
dark matter. Thus, the pieces of Eq. (\ref{61}) undergo the correspondences%
\begin{equation}
\nabla ^{\rho \text{ }\ast }R^{\lambda }{}_{[\rho \sigma ]\lambda }\mapsto
-i(\nabla _{C^{\prime }}^{B}\xi _{BC}{}-\text{c.c.})  \label{64}
\end{equation}%
and%
\begin{equation}
-2^{\ast }T^{\lambda \rho \tau }R_{\tau \lbrack \rho \sigma ]\lambda
}\mapsto -i(\varepsilon ^{DB}\tau ^{D^{\prime }B^{\prime }AA^{\prime }}-%
\text{c.c.})(\varepsilon _{B^{\prime }C^{\prime }}R_{AA^{\prime
}(BC)L^{\prime }}{}^{L^{\prime }}{}_{DD^{\prime }}+\text{c.c.}),  \label{70}
\end{equation}%
with the unprimed X-contributions carried by the $\varepsilon R$-pieces of (%
\ref{70}) being conveniently added together such that%
\begin{equation}
\text{X}_{A(BC)D}=\Psi _{ABCD}-\frac{1}{2}\varepsilon _{AD}\xi _{BC}+\frac{1%
}{6}\varkappa \varepsilon _{A(B}\varepsilon _{C)D}  \label{73}
\end{equation}%
and%
\begin{equation}
\text{X}_{LAD}{}^{L}=-(\xi _{AD}-\frac{1}{2}\varkappa \varepsilon _{AD}).
\label{74}
\end{equation}%
Equation (\ref{61}) therefore gets translated into%
\begin{equation}
\nabla _{C^{\prime }}^{B}\xi _{BC}{}+m_{CC^{\prime }}=\sigma _{CC^{\prime
}}^{(\text{X})}+\sigma _{CC^{\prime }}^{(\Xi )},  \label{91}
\end{equation}%
with%
\begin{equation}
m_{CC^{\prime }}=(T_{C^{\prime }}^{A}-3\tau ^{AB}{}_{BC^{\prime }})\xi _{AC}+%
\frac{3}{2}\tau ^{AB}{}_{CC^{\prime }}{}\xi _{AB}.  \label{91Lin}
\end{equation}

The $\sigma $-pieces of Eq. (\ref{91}) are sources for $\xi _{BC},$ which
are given by certain couplings between torsion kernels and $\varkappa \Psi
\Xi $-curvatures [25]. Such as in the description of the cosmic backgrounds,
the $m$-terms borne by the field equations for $\xi _{BC}{}$ and $\xi
_{B^{\prime }C^{\prime }}{}$ emerge strictly from the gravitational
interaction between torsion kernels and the fields themselves, while
likewise carrying the information on part of the masses acquired by the
fields. The information on the whole masses of the $\xi $-fields had been
attained with the help of the differential techniques supplied in Ref. [24]
in conjunction with the procedures for deriving the corresponding wave
equations and, in fact, turned out to be of the same form as those given by
Eqs. (\ref{d1}) and (\ref{d2}). In the geometric situation posed by the
condition (\ref{addLin1}), the wave equation for $\xi _{AB}$ comes up, then,
as the invariant statement [25]%
\begin{equation}
(\square +\mu _{DM}^{2}+\frac{4}{3}\varkappa )\xi _{AB}-2\Psi
_{AB}{}^{CD}\xi _{CD}=-2(\nabla _{(A}^{C^{\prime }}\sigma _{B)C^{\prime }}^{(%
\text{X})}+\nabla _{(A}^{C^{\prime }}\sigma _{B)C^{\prime }}^{(\Xi )}),
\label{101}
\end{equation}%
with the device%
\begin{equation}
\mu _{DM}^{2}\xi _{AB}=-4\nabla _{\mu }(\xi _{C(A}\tau _{B)}^{C\mu
})-2\nabla _{(A}^{C^{\prime }}m_{B)C^{\prime }}.  \label{102}
\end{equation}

\section{Dirac Particles}

In either formalism, a Dirac system can be defined as the conjugate field
pairs borne by the set 
\begin{equation}
\mathfrak{D}=\{(\zeta ^{A},\chi _{A^{\prime }}),(\chi _{A},\zeta ^{A^{\prime
}})\}.  \label{d4}
\end{equation}%
All fields of this set possess the same rest mass $M$ which is, then, taken
to be invariant under complex conjugation. The entries of each pair bear the
opposite helicity values $+1/2$ and $-1/2$, but such values get reversed
when we pass from one pair to the other. Additionally, the fields of either
pair carry the same electric charge, with the charge of one pair being
opposite to the charge of the other pair. In the $\varepsilon $-formalism,
the unprimed and primed elements of the former pair appear as gauge
invariant spin-vector densities of weight $+1/2$ and antiweight $-1/2,$
respectively, with the values of the weight and antiweight for the latter
pair being the other way about. Whereas the weight and antiweight for the
pair $(\zeta _{A},\chi ^{A^{\prime }})$ thus appear as $-1/2$ and $+1/2,$
the ones for $(\chi ^{A},\zeta _{A^{\prime }})$ are $+1/2$ and $-1/2,$ but
the electric charge of either pair of $\mathfrak{D}$ remains unaltered under
any lowering and raising of indices.\footnote{%
By definition, $M$ is also invariant under any lowering and raising of
indices.}

The gauge characterization of any $\varepsilon $-formalism Dirac wave
function matches some of the geometric situations described in Ref. [24],
which invariantly entail the absence of $\phi \psi $-contributions from
expansions that involve the differential operator%
\begin{equation}
\check{D}_{AB}=-\nabla _{(A}^{C^{\prime }}\nabla _{B)C^{\prime }}+2\tau
_{AB}{}^{\mu }\nabla _{\mu },  \label{35}
\end{equation}%
with%
\begin{equation}
D_{\mu \nu }\mapsto (\check{D}_{AB},\check{D}_{A^{\prime }B^{\prime }}).
\label{add15}
\end{equation}%
Towards obtaining the wave equations that could expectedly bring out the
couplings of Dirac fields with the cosmic backgrounds, we shall consequently
have to utilize an adequate minimal coupling covariant derivative operator
along with its conjugate. In what follows, we will utilize the
configurations exhibited in the foregoing Sections along with $\check{D}$%
-derivatives to elaborate upon the main aim of our work. We will carry out
the formulation of Dirac's theory for the pair $(\zeta ^{A},\chi _{A^{\prime
}}),$ in connection with our present purposes. One shall be able to derive
the field and wave equations for the other pairs by taking complex
conjugations and invoking the covariant constancy of the $\varepsilon $%
-spinors.

Instead of allowing for the Weyl gauge group, which yields [22]
uninteresting invariant derivatives of the form%
\begin{equation}
\nabla _{\mu }^{\prime }\zeta ^{\prime A}=\nabla _{\mu }\zeta ^{A},\text{ }%
\nabla _{\mu }^{\prime }\chi _{A^{\prime }}^{\prime }=\nabla _{\mu }\chi
_{A^{\prime }},  \label{d16}
\end{equation}%
we must implement non-geometric gauge transformations like\footnote{%
We remind that the natural system of units has been adopted by us.}%
\begin{equation}
\zeta ^{\prime A}=e^{-i\breve{e}\theta }\zeta ^{A},\text{ }\chi _{A^{\prime
}}^{\prime }=e^{-i\breve{e}\theta }\chi _{A^{\prime }},\text{ }\breve{A}%
_{\mu }^{\prime }=\breve{A}_{\mu }-\partial _{\mu }\theta ,  \label{d17}
\end{equation}%
where $\breve{e}$ stands for the charge of the field pair being considered, $%
\theta $ is a world-spin invariant, and $\breve{A}_{\mu }$ amounts to the
sum of the potentials that occur in Eq. (\ref{17}), that is to say,%
\begin{equation}
\breve{A}_{\mu }=\Phi _{\mu }+A_{\mu }.  \label{d30}
\end{equation}%
Hence, by defining the operator%
\begin{equation}
d_{\mu }=\nabla _{\mu }-i\breve{e}\breve{A}_{\mu },  \label{d15}
\end{equation}%
we obtain the covariant differential patterns%
\begin{equation}
d_{\mu }^{\prime }\zeta ^{\prime A}=e^{-i\breve{e}\theta }d_{\mu }\zeta ^{A}
\label{d50}
\end{equation}%
and%
\begin{equation}
d_{\mu }^{\prime }\chi _{A^{\prime }}^{\prime }=e^{-i\breve{e}\theta }d_{\mu
}\chi _{A^{\prime }},  \label{d51}
\end{equation}%
whence Dirac's field equations are invariantly stated as%
\begin{equation}
(\nabla _{BB^{\prime }}-i\breve{e}\breve{A}_{BB^{\prime }})\zeta ^{B}=-\mu
\chi _{B^{\prime }}  \label{d19}
\end{equation}%
and%
\begin{equation}
(\nabla ^{BB^{\prime }}-i\breve{e}\breve{A}^{BB^{\prime }})\chi _{B^{\prime
}}=\mu \zeta ^{B},  \label{d21}
\end{equation}%
where $\mu =M/\sqrt{2}$ is the normalized rest mass.

For deriving the relevant wave equations systematically, it is expedient to
introduce the operator%
\begin{equation}
\eth _{\mu \nu }=d_{\mu }d_{\nu },  \label{d25}
\end{equation}%
which, upon acting on any symbolic field quantity $\Upsilon $ of charge $%
\breve{e},$ gives the parts%
\begin{equation}
\eth _{[\mu \nu ]}\Upsilon =(\nabla _{\lbrack \mu }\nabla _{\nu ]}+i\breve{e}%
T_{\mu \nu }{}^{\lambda }\breve{A}_{\lambda }+\frac{1}{4}\breve{e}C_{\mu \nu
M}{}^{M})\Upsilon  \label{90}
\end{equation}%
and%
\begin{equation}
\eth _{(\mu \nu )}\Upsilon =(\nabla _{(\mu }\nabla _{\nu )}-2i\breve{e}%
\breve{A}_{(\mu }\nabla _{\nu )})\Upsilon -i\breve{e}(d_{(\mu }\breve{A}%
_{\nu )})\Upsilon ,  \label{99}
\end{equation}%
where Eqs. (\ref{16}) and (\ref{17}) have been employed together with the
equality%
\begin{equation}
i\breve{e}\nabla _{(\mu }\breve{A}_{\nu )}+\breve{e}^{2}\breve{A}_{\mu }%
\breve{A}_{\nu }=i\breve{e}d_{(\mu }\breve{A}_{\nu )}.  \label{e901}
\end{equation}%
Supposing that $\Upsilon ^{\prime }=e^{-i\breve{e}\theta }\Upsilon ,$ and
performing some straightforward computations, we thus establish the
covariant behaviours%
\begin{equation}
\eth _{[\mu \nu ]}^{\prime }\Upsilon ^{\prime }=e^{-i\breve{e}\theta }\eth
_{[\mu \nu ]}\Upsilon ,\text{ }\eth _{(\mu \nu )}^{\prime }\Upsilon ^{\prime
}=e^{-i\breve{e}\theta }\eth _{(\mu \nu )}\Upsilon .  \label{behav1}
\end{equation}%
If the relations (\ref{26}) and (\ref{26Lin}) are called upon together with
Eq. (\ref{35}), we may then spell out the following derivative expressions
for $\zeta ^{C}$%
\begin{equation}
\eth _{(AB)A^{\prime }}^{A^{\prime }}\zeta ^{C}=[-\check{D}_{AB}+2\tau
_{AB}{}^{\mu }\nabla _{\mu }+i\breve{e}(\phi _{AB}+\psi _{AB}-2\tau
_{AB}{}^{\mu }\breve{A}_{\mu })]\zeta ^{C}  \label{501}
\end{equation}%
and%
\begin{equation}
\eth _{[AB]A^{\prime }}^{A^{\prime }}\zeta ^{C}=\frac{1}{2}\varepsilon
_{AB}[(2i\breve{e}\breve{A}^{\mu }\nabla _{\mu }-\square )\zeta ^{C}+(i%
\breve{e}d_{\mu }\breve{A}^{\mu })\zeta ^{C}],  \label{d503}
\end{equation}%
with Eq. (\ref{d503}) yielding the relation%
\begin{equation}
\eth _{\mu }{}^{\mu }\zeta ^{C}=(\square -2i\breve{e}\breve{A}^{\mu }\nabla
_{\mu })\zeta ^{C}-(i\breve{e}d_{\mu }\breve{A}^{\mu })\zeta ^{C},
\label{502}
\end{equation}%
which agrees with (\ref{99}). So, by contracting the indices $B$ and $C$ of
the configurations (\ref{501}) and (\ref{d503}), likewise invoking Eqs. (\ref%
{d19}) and (\ref{d21}), we write the wave equation%
\begin{equation}
\eth ^{AA^{\prime }}{}_{BA^{\prime }}\zeta ^{B}+\mu ^{2}\zeta ^{A}=0,
\label{w1}
\end{equation}%
which, in view of (\ref{behav1}), behaves covariantly under the
transformations (\ref{d17}), namely,%
\begin{equation}
\eth ^{\prime AA^{\prime }}{}_{BA^{\prime }}\zeta ^{\prime B}+\mu ^{2}\zeta
^{\prime A}=e^{-i\breve{e}\theta }(\eth ^{AA^{\prime }}{}_{BA^{\prime
}}\zeta ^{B}+\mu ^{2}\zeta ^{A}).  \label{g1}
\end{equation}

Because $\zeta ^{A}$ is a contravariant one-index spin tensor density of
weight $1/2,$ the $\check{D}$-derivative that occurs in Eq. (\ref{501})
amounts to the gravitational configuration%
\begin{equation}
\check{D}_{AB}\zeta ^{C}=\text{X}_{ABD}{}^{C}\zeta ^{D},  \label{add903}
\end{equation}%
which produces the $\Psi $-free expansion%
\begin{equation}
\check{D}_{AB}\zeta ^{B}=-\xi _{AB}\zeta ^{B}-\frac{1}{2}\varkappa \zeta
^{B}\varepsilon _{BA},  \label{add502}
\end{equation}%
since the coupling $\Psi _{ABD}{}^{C}\zeta ^{D}$ gets annihilated when the
indices $B$ and $C$ are contracted. The $\tau \nabla $-piece of (\ref{501})
can be naively worked out [25] by taking account of the geometric
circumstance stipulated by Eq. (\ref{addLin1}), in which case we can write
the particular device%
\begin{equation}
\tau _{AB}{}^{\mu }\nabla _{\mu }\zeta ^{C}=\nabla _{\mu }(\tau _{AB}{}^{\mu
}\zeta ^{C})-\xi _{AB}\zeta ^{C},  \label{add507}
\end{equation}%
in a way similar to (\ref{add100}). Equation (\ref{w1}) can therefore be
written out explicitly when the configurations (\ref{501})-(\ref{add507})
are combined together. We have, in effect,%
\begin{equation}
(\square +2\mu ^{2}+\varkappa )\zeta ^{A}+4\nabla _{\mu }(\tau
_{B}^{A}{}^{\mu }\zeta ^{B})+2i\breve{e}(\phi _{B}^{A}+\psi _{B}^{A})\zeta
^{B}-2\xi _{B}^{A}\zeta ^{B}=0,  \label{531}
\end{equation}%
along with the subsidiary relation%
\begin{equation}
i\breve{e}\breve{A}^{\mu }(\nabla _{\mu }\zeta ^{A}+2\tau _{B\mu
}^{A}{}\zeta ^{B})+\frac{1}{2}i\breve{e}(d_{\mu }\breve{A}^{\mu })\zeta
^{A}=0,  \label{cond1}
\end{equation}%
which gives rise to the coupled gauge conditions%
\begin{equation}
i\breve{e}\nabla ^{\mu }\theta (\nabla _{\mu }\zeta ^{A}+2\tau _{B\mu
}^{A}{}\zeta ^{B})=0  \label{gauge1}
\end{equation}%
and\footnote{%
Equation (\ref{gauge1}) enters the description as a gauge condition for $%
\zeta ^{A}$ while Eq. (\ref{gauge2}) amounts to a gauge condition for $%
\breve{A}^{\mu }.$ These conditions assure the gauge invariance of the
relation (\ref{cond1}).}%
\begin{equation}
\square \theta -i\breve{e}\nabla ^{\mu }\theta (\nabla _{\mu }\theta )=0.
\label{gauge2}
\end{equation}

For $\chi _{A^{\prime }},$ we have the invariant wave equation%
\begin{equation}
\eth {}_{AA^{\prime }}{}^{AB^{\prime }}\chi _{B^{\prime }}+\mu ^{2}\chi
_{A^{\prime }}=0,  \label{w2}
\end{equation}%
as well as the formulae%
\begin{equation}
\eth _{A}^{(A^{\prime }B^{\prime })A}{}\chi _{C^{\prime }}=[\check{D}%
^{A^{\prime }B^{\prime }}-2\tau ^{A^{\prime }B^{\prime }\mu }\nabla _{\mu }-i%
\breve{e}(\phi ^{A^{\prime }B^{\prime }}+\psi ^{A^{\prime }B^{\prime
}}-2\tau ^{A^{\prime }B^{\prime }\mu }\breve{A}_{\mu })]\chi _{C^{\prime }}
\label{503}
\end{equation}%
and%
\begin{equation}
\eth _{A}^{[A^{\prime }B^{\prime }]A}{}\chi _{C^{\prime }}=\frac{1}{2}%
\varepsilon ^{A^{\prime }B^{\prime }}[(\square -2i\breve{e}\breve{A}^{\mu
}\nabla _{\mu })\chi _{C^{\prime }}-(i\breve{e}d_{\mu }\breve{A}^{\mu })\chi
_{C^{\prime }}],  \label{504}
\end{equation}%
with Eq. (\ref{504}) implying that%
\begin{equation}
\eth _{\mu }{}^{\mu }\chi _{C^{\prime }}=(\square -2i\breve{e}\breve{A}^{\mu
}\nabla _{\mu })\chi _{C^{\prime }}-(i\breve{e}d_{\mu }\breve{A}^{\mu })\chi
_{C^{\prime }}.  \label{507}
\end{equation}%
Hence, by using the $\chi $-version of (\ref{add507}) together with the
contracted derivative%
\begin{equation}
\check{D}^{A^{\prime }B^{\prime }}\chi _{B^{\prime }}=-\xi ^{A^{\prime
}B^{\prime }}\chi _{B^{\prime }}+\frac{1}{2}\overline{\varkappa }\varepsilon
^{A^{\prime }B^{\prime }}\chi _{B^{\prime }},  \label{add504}
\end{equation}%
and fitting pieces together afterwards, we arrive at the wave equation%
\begin{equation}
(\square +2\mu ^{2}+\overline{\varkappa })\chi _{A^{\prime }}-4\nabla _{\mu
}(\tau _{A^{\prime }}^{B^{\prime }}{}^{\mu }\chi _{B^{\prime }})-2i\breve{e}%
(\phi _{A^{\prime }}^{B^{\prime }}+\psi _{A^{\prime }}^{B^{\prime }})\chi
_{B^{\prime }}+2\xi _{A^{\prime }}^{B^{\prime }}\chi _{B^{\prime }}=0,
\label{532}
\end{equation}%
which must be constrained to the relation%
\begin{equation}
i\breve{e}\breve{A}^{\mu }(\nabla _{\mu }\chi _{A^{\prime }}-2\tau
_{A^{\prime }\mu }^{B^{\prime }}{}\chi _{B^{\prime }})+\frac{1}{2}i\breve{e}%
(d_{\mu }\breve{A}^{\mu })\chi _{A^{\prime }}=0.  \label{cond2}
\end{equation}%
Evidently, Eq. (\ref{cond2}) reinstates the pattern (\ref{gauge2}) whilst
supplying the $\chi $-condition%
\begin{equation}
i\breve{e}\nabla ^{\mu }\theta (\nabla _{\mu }\chi _{A^{\prime }}-2\tau
_{A^{\prime }\mu }^{B^{\prime }}{}\chi _{B^{\prime }})=0,  \label{gauge3}
\end{equation}%
whence it bears invariance under the transformations (\ref{d17}).

It should be emphasized that the above-derived wave equations display only
the couplings between the cosmic backgrounds and the $\zeta \chi $-fields
allowed for, which come about as a consequence of the implementation of the
operator $d_{\mu }.$ Of course, the adoption of the relations (\ref{cond1})
and (\ref{cond2}) must be aggregated into the formulation of Eqs. (\ref{531}%
) and (\ref{532}), while the choices (\ref{gauge1}), (\ref{gauge2}) and (\ref%
{gauge3}) guarantee that the invariance of Eqs. (\ref{w1}) and (\ref{w2})
under (\ref{d17}) may be carried over to (\ref{531}) and (\ref{532}).

\section{Concluding Remarks and Outlook}

The descriptions we have presented in Sections 3 and 4 tell us that it might
seem suggestive to say that any Dirac field had acquired an effective mass
soon after the occurrence of the big-bang just on an equal interacting
footing to the cosmic backgrounds and dark matter. Indeed, the wave
equations (\ref{531}) and (\ref{532}) may be reinstated so as to carry the
contributions%
\begin{equation}
m_{\zeta }^{2}\zeta ^{A}=2\mu ^{2}\zeta ^{A}+4\nabla _{\mu }(\tau
_{B}^{A}{}^{\mu }\zeta ^{B})  \label{cr1}
\end{equation}%
and%
\begin{equation}
m_{\chi }^{2}\chi _{A^{\prime }}=2\mu ^{2}\chi _{A^{\prime }}-4\nabla _{\mu
}(\tau _{A^{\prime }}^{B^{\prime }}{}^{\mu }\chi _{B^{\prime }}).
\label{cr1lin}
\end{equation}%
Since every one-index spin object bears an invariant nullity, it becomes
impossible to deduce from these contributions any symbolic expression for
either of $m_{\zeta }$ and $m_{\chi }$ that does not mix up charges. For $%
m_{\zeta },$ for instance, it follows that the uncharged expression%
\begin{equation}
m_{\zeta }=\sqrt{\frac{[2\mu ^{2}\zeta ^{A}+4\nabla _{\mu }(\tau
_{B}^{A}{}^{\mu }\zeta ^{B})]\chi _{A}}{\zeta ^{B}\chi _{B}}},  \label{cr2}
\end{equation}%
along with the reality requirement $m_{\zeta }^{2}>0,$ can be used for
carrying out a direct estimation procedure. Hence, the $\zeta \chi $%
-effective masses will bear invariance under the transformations (\ref{d17})
if the orthogonality conditions%
\begin{equation}
i\breve{e}\tau _{A}^{B}{}^{\mu }\zeta ^{A}\chi _{B}\nabla _{\mu }\theta =0
\label{cr5}
\end{equation}%
and%
\begin{equation}
i\breve{e}\tau _{B^{\prime }}^{A^{\prime }}{}^{\mu }\chi _{A^{\prime }}\zeta
^{B^{\prime }}\nabla _{\mu }\theta =0,  \label{cr5Lin}
\end{equation}%
are taken into account. Therefore, the gauge choices (\ref{gauge1}) and (\ref%
{gauge3}) may ultimately be simplified to%
\begin{equation}
i\breve{e}\nabla ^{\mu }\theta (\chi _{A}\nabla _{\mu }\zeta ^{A})=0
\label{cr9}
\end{equation}%
and%
\begin{equation}
i\breve{e}\nabla ^{\mu }\theta (\zeta ^{A^{\prime }}\nabla _{\mu }\chi
_{A^{\prime }})=0.  \label{cr9Lin}
\end{equation}

It is the expansion (\ref{add6}) which points out that gravitons and dark
matter were produced together by a torsional big-bang. Whence, as remarked
in Ref. [25], it may be said that the earliest states of very high density
of spinning matter must have constituted a non-conformally flat regime of
the cosmic evolution, regardless of whether account of Eqs. (\ref{add100})
and (\ref{add507}) is taken or not. The absence from the derivatives (\ref%
{add502}) and (\ref{add504}) of couplings between gravitons and Dirac
particles nevertheless allows us to assert that the propagation of quarks
and leptons was not affected in any way by any later occurrence of vanishing
wave functions for gravitons. Only macroscopically do Dirac particles
interact with gravitational fields as wave functions for gravitons carry a
microscopic character even at a classical level. We can conclude that, if
the framework we have developed actually bears a sufficient degree of
elementarity, then one could expect that gravitons may enter a unified
interaction model of particle physics only through couplings which are
equivalent to the uncharged bosonic patterns%
\[
\Psi {}_{AB}{}^{CD}B_{CD},\text{ }\Psi _{MN(AB}{}\Psi {}_{CD)}{}^{MN}. 
\]

One of the noteworthy outcomes we have obtained, which holds up a
considerable deal of importance, is the accomplishment of a theoretical
indication whereby dark energy and dark matter must have taken part in a
process that somehow carried forward the early spacetime torsionfulness to
the interior of hadrons during the nucleosynthesis era, through couplings
that typically look like%
\[
\xi _{(A}^{C}\psi _{B)C},\text{ }\breve{e}\psi _{B}^{A}\zeta ^{B},\text{ }%
\xi _{B}^{A}\zeta ^{B}. 
\]%
Remarkably enough, this indication conforms to the fact that
Einstein-Cartan's theory identifies just as dark matter the densities of
spinning matter which bring about the local production of spacetime torsion.
At least in principle, our field and wave equations can not recover the
standard magnitudes of the coupling constants for the interactions between
hadronic constituents. However, while attributing by means of the
transformations (\ref{d17}) an intrinsically non-geometric character to
Dirac's theory, our work has exhibited a formal evidence that dark bosons
had partaken of a torsion-preserving hadronic formation. Thus, from our
point of view, the dynamics which takes place in the interior of hadrons
must have kept the cosmic torsionfulness during the occurrence of
hadronization, in such a manner that both of $\psi _{AB}$ and $\xi _{AB}$
could bear stability as they interact inside the hadronic environment via $%
\xi _{(A}^{C}\psi _{B)C}.$ On the other hand, the couplings $\breve{e}\psi
_{B}^{A}\zeta ^{B}$ and $\xi _{B}^{A}\zeta ^{B}$ could eventually be
regarded as states for composite heavy charged fermions. It might be
expected, also, that both dark bosons would rapidly fade away whenever there
occurs a process which produces a loss of environmental spacetime torsion.
Under these conditions, the torsionless index-pair symmetry X$_{ABCD}=$ X$%
_{CDAB}$ would start holding, whilst $\psi _{AB}$ and $\xi _{AB}$ would both
turn out to vanish identically. As we believe, the Einstein-Cartan
ascription to dark matter of the spinning matter that locally produces
spacetime torsion, may likewise enable one to construct a complete scenario
of the very early stages of the evolution of the Universe, which involves
the concept of torsion at a microscopic level.

\end{document}